\documentclass[amsmath,amssymb,showkeys,nofootinbib]{revtex4}
\usepackage{graphicx}
\usepackage{subfig}
\usepackage{graphics}
\usepackage{amsfonts}
\usepackage{mathrsfs}
\usepackage{lscape}
\usepackage{amsmath,dcolumn,epsfig,rotating,amssymb,bm}
\usepackage{dcolumn}% Align table columns on decimal point
\usepackage{bm}% bold math

\begin{document}

%\title{A brand new robust method for distinguishing vertebrates from invertebrates using only a random, tiny, fraction of their miDNA sequences}
\title{A new method for identifying vertebrates using only their mitochondrial DNA}
\author{Nikesh S. Dattani}
%\vspace{10mm} doesn't do anything!
%\affiliation{Department of Computer Science \\ University of Western Ontario, London, Ontario~ \\  N6A 3K7, Canada}
\affiliation{Department of Materials, Oxford University, OX1 3PH, United Kingdom}
\date{\today}
\begin{abstract}
A new method for determining whether or not a mitrochondrial DNA (mtDNA) sequence belongs to a vertebrate is described and tested. This method only needs the mtDNA sequence of the organism in question, and unlike alignment based methods, it does not require it to be compared with anything else. The method is tested on all 1877 mtDNA sequences that were on NCBI's nucleotide database on August 12, 2009, and works in 94.57\% of the cases. Furthermore, all organisms on which this method failed are closely related phylogenetically in comparison to all other organisms included in the study. A list of potential extensions to this method and open problems that emerge out of this study is presented at the end.
\end{abstract}

\keywords{mtDNA, nucleotide frequency, vertebrata}

\maketitle

%\begin{center}
%Department of Computer Science, University of Western Ontario \\ London, Ontario~
% N6A 3K7, Canada
%\end{center}

\vspace{10mm}

\section{Preliminary Observations}

The definition of the mononucleotide frequency matrix for a DNA sequence $\sigma$ is:

\begin{equation*}
F_1 (\sigma) =
\begin{bmatrix}
N_C & N_G \\
N_A & N_T \\
\end{bmatrix},
\label{eq:matrix}  
\end{equation*}

\noindent
where $N_X$ represents the number of occurrences of nucleotide $X$.
%  This matrix is often used in studying the `frequency chaos game representation' (FCGR) of  DNA sequences\cite{99desc,01alme,05wang}
Four examples of such matrices are presented below -- each representing the complete mtDNA sequence for a different organism:

\begin{equation*}
F_1 (a) =
\begin{bmatrix}
1949 & 1481 \\
6326 & 6263 \\
\end{bmatrix} ~~,~~
F_1 (b) =
\begin{bmatrix}
3523 & 2600 \\
4353 & 4282 \\
\end{bmatrix},
\vspace{-4mm}
\end{equation*}

%\hspace{100mm} %%%%%%%%%%%%% GET THIS TO WORK!!!
\begin{equation*}
F_1 (c) =
\begin{bmatrix}
4740 & 2590 \\
4954 & 4254 \\
\end{bmatrix} ~~,~~
F_1 (d) =
\begin{bmatrix}
5159 & 2201 \\
5605 & 4778 \\
\end{bmatrix},
\end{equation*}

\noindent
{\small
$a =$ \emph{Drosophila yakuba} (a fruit fly),\\
$b =$ \emph{Chlamydomonas reinhardtii} (a unicellular green alga),\\
$c =$ \emph{Abalistes stellaris} (starry triggerfish),\\
$d =$ \emph{Polychrus marmoratus} (a many-coloured lizard).\\ }

\noindent
It is immediately apparent that in the former two cases, $N_C \approx N_G$ and $N_A \approx N_T$, while in the latter two, $N_C \approx N_T$ and $N_T \approx N_A$.  The similarities in these permutations of the four nucleotides are quite striking when compared to their highly dissimilar counterparts (take for example, the difference between $N_G$ and $N_A$ in sequence $a$!).

Furthermore, examining many more cases leads me to believe that there is a salient resemblance between the above two categories and accepted taxonomical clades.

In order to test this hypothesis, I have designed a metric to quantify how `close' a sequence's nucleotide frequencies are for any 2-set of the alphabet $\{C,G,A,T\}$.  I define this metric the \emph{mononucleotide frequency similarity measure}:

\begin{equation}
\mathscr{D}_{XY} = (N_X - N_Y)^{-2},
\label{eq:metric}
\end{equation}

\noindent
where $X$ and $Y$ represent any nucleotide.  The reason for the negative exponent was inspired by my desire to isolate \emph{small} differences (the closer together two mononucleotide frequencies are, the larger the measure should be), and the squaring is to ensure that the measure does not change sign.

\begin{figure*}[h!]
   \centering
\includegraphics[scale=0.4]{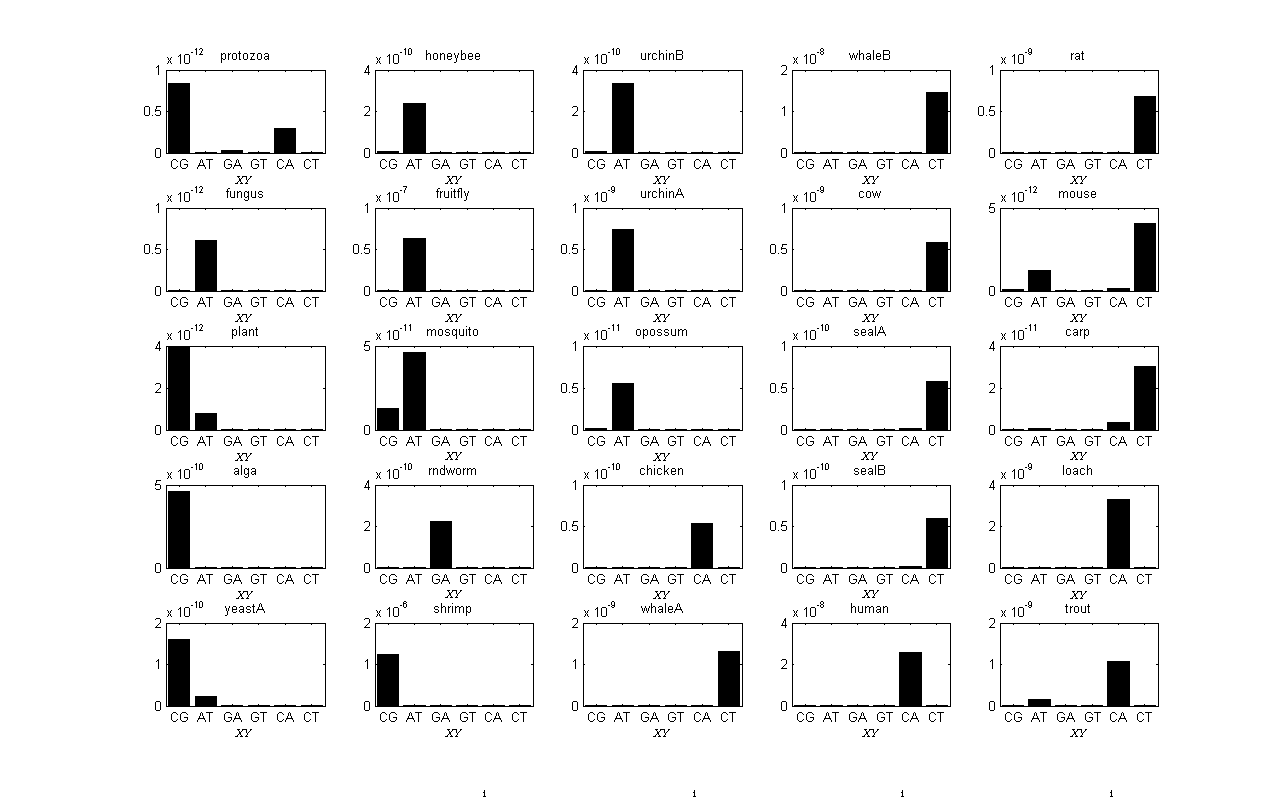}
	\caption{A preliminary test of the mononucleotide frequency similarity metric for the set of organisms studied in ref. \cite{2005Wang}. All vertebrates (except for one) have their highest peak in one of the rightmost bins, and all others have their first peak in one of the leftmost bins.}   
   \label{fig:yingwei}
 \end{figure*}

This metric is tested on a set of organisms below. To avoid any bias in the choices for these organisms, I chose the same set of organisms that were used in a completely different study (see ref. \cite{2005Wang}). I presented them in the exact order that they appear in fig. 2a of that paper (reproduced as fig. \ref{fig:yingweiPhylo} below), and removed one organism\footnote{I originally did this to make the figure 5x5, but I may change it to 13x2 for the next version.}. According to the aforementioned hypothesis, it is expected that $\mathscr{D}_{CT}$ and $\mathscr{D}_{TA}$ should dominate the rest of the $\mathscr{D}_{XY}$ values in the case of the vertebrates, while $\mathscr{D}_{CG}$ and $\mathscr{D}_{AT}$ should dominate in all other cases -- therefore, in the diagram below, it is expected that the vertebrates have sharp peaks in either of the two rightmost bins, and for the non-vertebrates to have sharp peaks in either of the two leftmost bins. Henceforth, the spectra in fig. \ref{fig:yingwei} will be referred to as the $\mathscr{D}$-spectra of the organisms in question, and the terms `first peak' and `second peak' will respectively refer to the highest peak and second highest peaks in a particular $\mathscr{D}$-spectrum.

\begin{figure*}[h!]
   \centering
\includegraphics[scale=0.5]{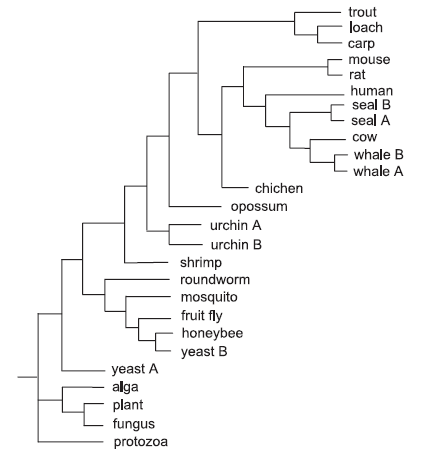}
	\caption{Fig. 2a of ref. \cite{2005Wang}. The way in which this phylogenetic tree was calculated is explained in \cite{2005Wang}, and this phylogenetic calculation resulted in the opposum being the closest vertebrate to the non-vertebrates out of all the organisms in that study. The opposum also happens to be the only vertebrate that refutes my conjecture above.}   
   \label{fig:yingweiPhylo}
 \end{figure*}   

%If you keep the commented line below (which I originally put in the figure caption, it might become apparent to people that opposum being non-vertebrate like isn't such a special thing... 's just the consequence of the FCGR10

%This phylogenetic tree was generated by PHYLIP where the `distances' between each organism were defined as the Euclidean distances between their respective decanucleotide frequency matrices (analogous to the matrix of eq. \eqref{eq:matrix}, except with $2^10$x$2^10$ elements, each corresponding to the number of occurrences of a particular nucleotide of length 10).

The fact that the order in which these organisms are presented in fig. \ref{fig:yingwei} was unaltered from that of fig. 2a of ref. \cite{2005Wang} leads to another interesting observation: The only vertebrate in this dataset with its first peak in either of the leftmost bins is the opossum, and according to that phylogenetic tree, the opposum is the most closely related vertebrate in this data set to the non-vertebrates (notice its position in fig. \ref{fig:yingwei}).  This could potentially mean that the opossum and the non-vertebrates in this dataset belong to a clade separate from the other vertebrates - one where the presence of a backbone does not play a r\^{o}le.  This may also mean that when used for vertebrates, the mononucleotide frequency similarity measure is sensitive in some way to the organism's size (notice that the opossum is on average the smallest vertebrate in this dataset apart from the mouse, and the mouse's spectrum in fig. \ref{fig:yingwei} is the least in-focus of all the vertebrates).  In contrast, the opossum's $\mathscr{D}$-spectrum may simply be an anomaly that was also absorbed into the phylogenetic study of \cite{2005Wang}, thereby playing a vital r\^{o}le in its placement in that tree.

A second interesting feature in fig. \ref{fig:yingwei} is the $\mathscr{D}$-spectrum of the roundworm.  Its first peak appears in a matchless position when compared to the rest of the organisms in the dataset. As demonstrated in the supplementary document, this position for the first peak is relatively rare compared to the other four seen in fig. \ref{fig:yingwei}, and many of the organisms that share this property are related by accepted taxonomical classifications. 	

%Additionally, it may seem as though the order that these organisms appear in \ref{fig:yingwei} was chosen to emphasize this method's capability to categorize organisms into vertebrates and non-vertebrates, but in fact the order was unaltered from that of fig. 2a in \cite{05wang}.

%\subsection{Extensions}
%
%In order to improve the focus in the peaks of the $\mathscr{D}$-spectra, we may raise the metric \ref{eq:metric} by a positive exponent, thereby pressing down the more insignificant bars:
%
%\begin{equation}
%\mathscr{D}^{\, n}_{XY} = (N_X - N_Y)^{-n}, n>0.
%\label{eq:metric}
%\end{equation}
%
%\noindent
%For reasons explained in the supplementary data, we have chosen to use $n=4$.

\subsubsection{Secondary peaks}

\begin{figure*}[h!]
   \centering
\includegraphics[scale=0.4]{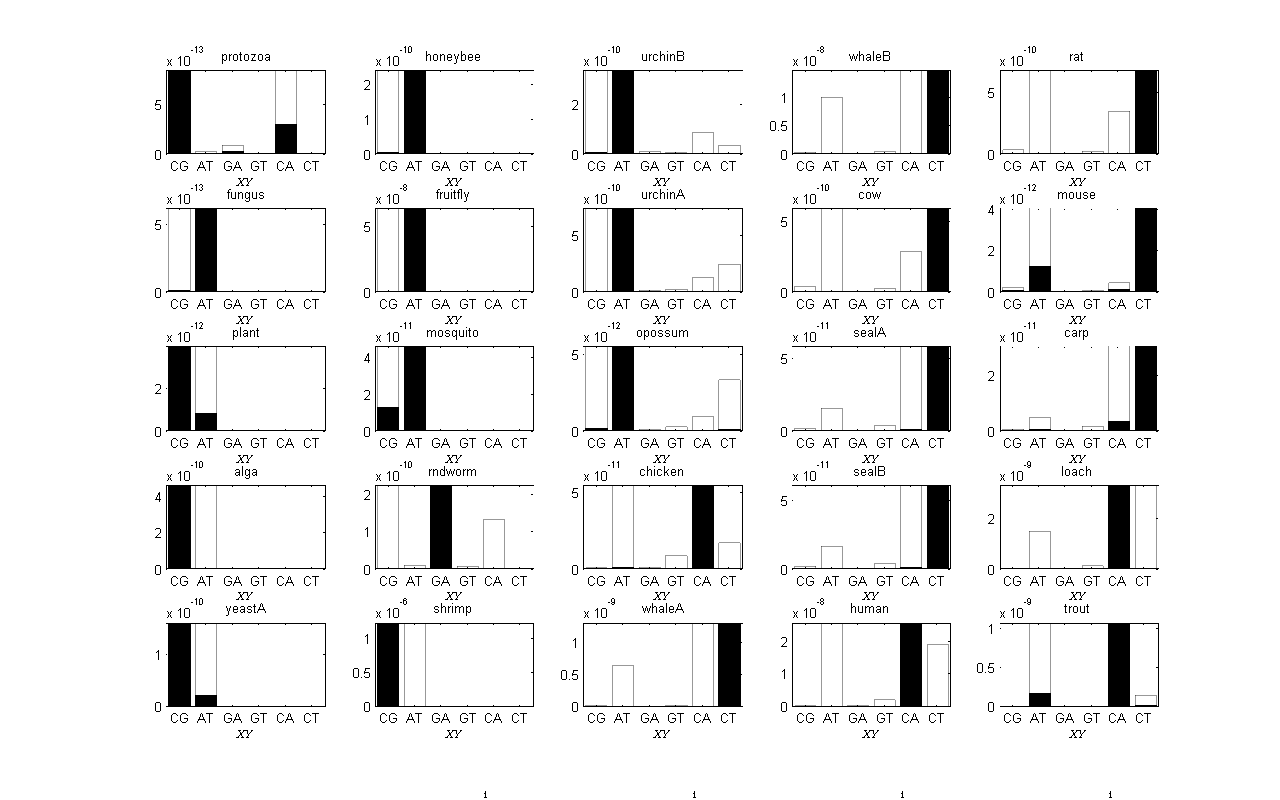}
	\caption{A replica of fig. \ref{fig:yingwei} except with the second peaks revealed.}   
   \label{fig:YWsecondPeaks}
 \end{figure*}

The black bars in fig. \ref{fig:YWsecondPeaks} below represent the exact same quantities as in fig. \ref{fig:yingwei}, except that they are all scaled by an appropriate factor in order for all of the first peaks to fill the vertical space available. The added white bars are the same as the black ones, except that the first peak is removed and the rest of them are scaled so that the second peak is as high as the first peak used to be. This is to reveal the locations of the second peaks.

Except for the protozoan, the second peaks of all non-vertebrates (including the roundworm, which appeared anomalous in fig. \ref{fig:yingwei}) are once again in one of the two left-most bins. As for the vertebrates, with the exception of the anomalous opossum, all secondary peaks are in either one of the right-most bins or at position AT. \textit{No first or second peaks arise at position GT} -- a facet that remains true in almost every organism in the NCBI nucleotide database (see supplementary document). 

\section{Benchmark results}

The data set is now extended to include \textit{all} 1877 mtDNA sequences found on the NCBI gene bank (1196 vertebrates and 681 non-vertebrates). The $\mathscr{D}$-sectra of each organism is available in the supplementary document. For the sake of argument, I will assume that the statistics of the $\mathscr{D}$-spectra of these 1877 organisms on NCBI are representative of all organisms with mitochondria. This assumption is not desirable, but it can't be avoided if we wish to continue and investigate the predictions below. As more organisms enter the NCBI genebank, the predictions below can be experimentally tested.

 One preliminary observation is that, out of all vertebrates, the highest peak was in the set \{CG, GA, GT\} only five times (0.41\% of all vertebrates). All five of these organisms were fish, and it is not surprising that out of all vertebrates, fish are the evolutionarily the closest to the invertebrates. If one were to find an unidentified mtDNA sequence with its highest peak either at any one of these three positions, they may be inclined to predict that the sequence does not belong to a vertebrate. Since only 5 of 1196 vertebrates seem to disobey this rule, and with the above assumption of consistency in mind, this prediction would be accurate 99.58\% of the time. 
 
 However, we can do much better. If we not only look at an organism's \textit{highest} peak in the $\mathscr{D}$-spectra, but the \textit{second highest} peak as well, we find that absolutely none of the 5 anomalies above have their second highest peak in the same set \{CG, GA, GT\}. This means that, if one were two find an unidentified mtDNA sequence with its highest two peaks being in the set \{CG, GA, GT\}, they may predict that the sequence does not belong to a vertebrate, and out of all mtDNA sequences available on NCBI so far, this prediction would be successful 100\% of the time.
 
 As exciting as the above observation may seem as a method to detect non-vertebrates, only 94 out of the 681 non-vertebrates (13.80\%) had the feature of having their first and second highest peaks satisfying the above criteria. So although 100\% accurate, the above prediction can only be applied to mtDNA sequences whose $\mathscr{D}$-spectra have a specific property which is not very common. 
 
 %The case where the highest peak is \textit{not} in the set \{CG, GA, GT\} will be addressed below, but after some more preliminary observations about the non-vertebrates are discussed.
 
 %Out of all non-vertebrates, the highest peak was in the set \{GT, CA, CT\} only 40 times (5.87\% of all non-vertebrates). 
 
 In order to set up a scheme in which vertebrates can be picked out faithfully, which is also applicable for a substantial percentage of cases, one needs to find a set of pairs of nucleotides for which either the vertebrates or the non-vertebrates never (or rarely ever) have their first or second peaks at positions corresponding to any pair of nucleotides in that set (as in the above case for vertebrates), while there does in fact exist plenty of organisms of the opposite clade that have peaks at those positions. 
 
 Such a scheme exists, and is based on the logic that follows. Only 42 non-vertebrates have their first peak in the set \{AT,CA,CT\} \textit{and} their second peak in the set \{AT,GT,CA,CT\}, while 1136 vertebrates do. So out of the 1178 total organisms with this property, about 96.4\% of them are vertebrates. The 1178 organisms with this property constitute about 62.8\% of the 1877 organisms taking part in the entire study, so this prediction can be made fairly often. Of the 699 organisms in the study that did not exhibit the above property, only 60 of them were vertebrates, and 639 of them were not. This means that for the remaining 37.2\% of the cases, about 90.1\% of them are not vertebrates.
 
 In light of the above observations, if one finds an unidentified mtDNA sequence, and the first two $\mathscr{D}$-spectrum peaks fall into category 1, he or she may predict that the organism belongs to the vertebrata clade. Under the above assumption of consistency, this will happen about 62.8\% of the time and will be accurate in 96.4\% of those cases. Otherwise (37.2\% of the time), he or she may predict that the organism does not belong to the vertebrata clade, and will be correct in 90.1\% of those cases. So in total, the probability that the investigator's prediction will be correct is 94.57\% (0.964x37.2\% + 90.1x37.2\% = $\frac{1136 + 699}{1877}$).
 
 The number of possible schemes analogous to the above is more than 10$^{1098}$ times the number of particles that we believe exist within the observable universe (10$^{90}$)\footnote{The number of different subsets that can be chosen for the first peaks is 2$^6$-1. For each of these subsets, one may choose from a list of 2$^6$-1 subsets corresponding to the second peaks, giving 3969 different combinations. Now the number of possible schemes that can be applied to identify the organism's clade is the number of proper subsets of those 3969 sets, which is 2$^{3969}$-1. Since 2$^{10} > 10^3$, $2^{3969} > 10^{1188} = 10^{90}10^{1098}$}. It would be impossible to search all of these possibilities to find the best scheme. The algorithm which I used to isolate the above scheme (which I do not claim is the best one possible), is explained in the supplementary data.
 
 % it's interesting to see based on the vertebrate data, what how often you'd be right if you did the opposite. If the highest peak was NOT in {CG,GA,GT}^2 , then what are the chances of it being a vertebrate, this involves looking at the number of nonvertebrates with this property.

\section{Possible Extensions}

A natural extension of the above analysis would be to investigate whether these peaks in the $\mathscr{D}$-spectra are accumulated erratically over the entire mitochondrial genome, or alternatively, whether this phenomenon occurs uniformly and consistently along the genome. If the latter is true, one would not need the entire mtDNA sequence in order to classify its origin. As a preliminary test, each genome in the data set of figs.\ref{fig:yingwei} \& \ref{fig:yingweiPhylo} was sliced into equal-sized pieces, and the $\mathscr{D}$-spectra of each slice was computed.  The results after slicing the genomes into quarters and eigths are displayed in figs. \ref{fig:3d4ths} \& \ref{fig:3d8ths} below, where the indices ascend according to the placement of their corresponding slice in the original fasta file.  The spectrum for each slice is also normalized to lie between 0 and 1 in order to keep the peaks of each slice visible.  Results for slicing the genome into different numbers of sequences, and unnormalized versions of the same results are provided in the supplementary documentation as well as results for all other organisms in this study.

\begin{figure}[h!]
   \centering
\includegraphics[scale=0.4]{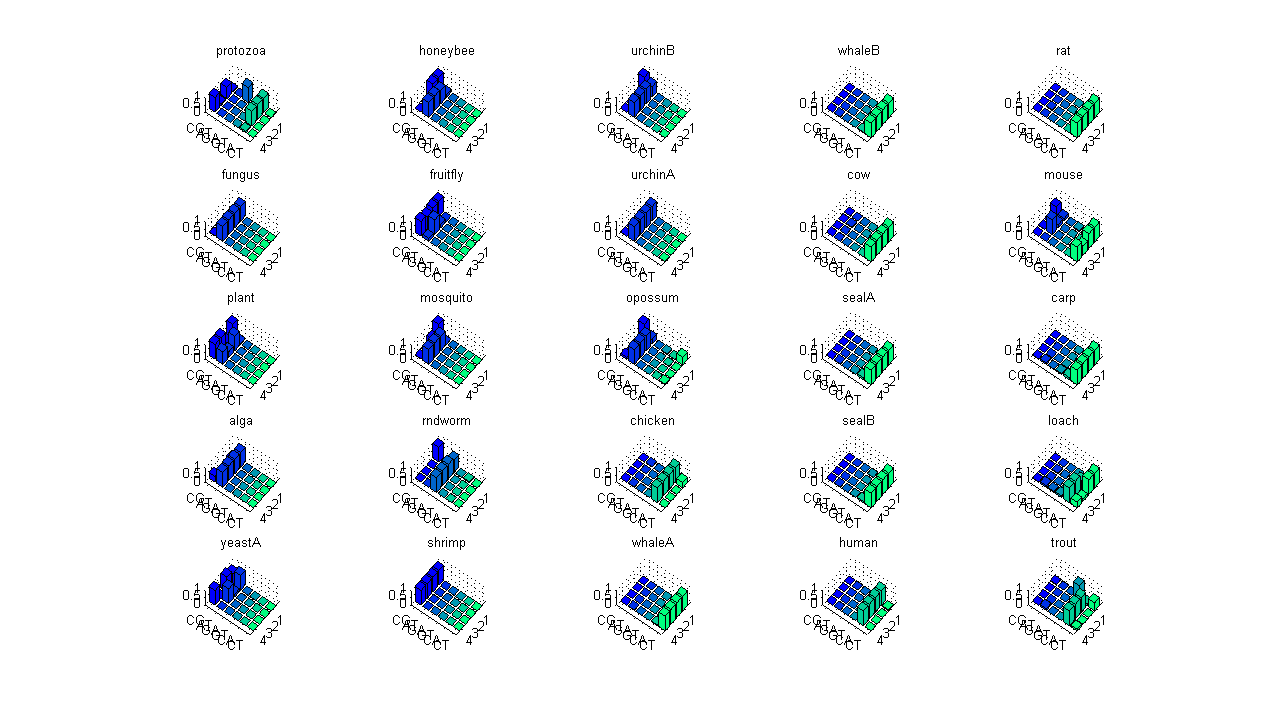}
	\caption{quarters}   
   \label{fig:3d4ths}
 \end{figure}   

\begin{figure}[h!]
   \centering
\includegraphics[scale=0.4]{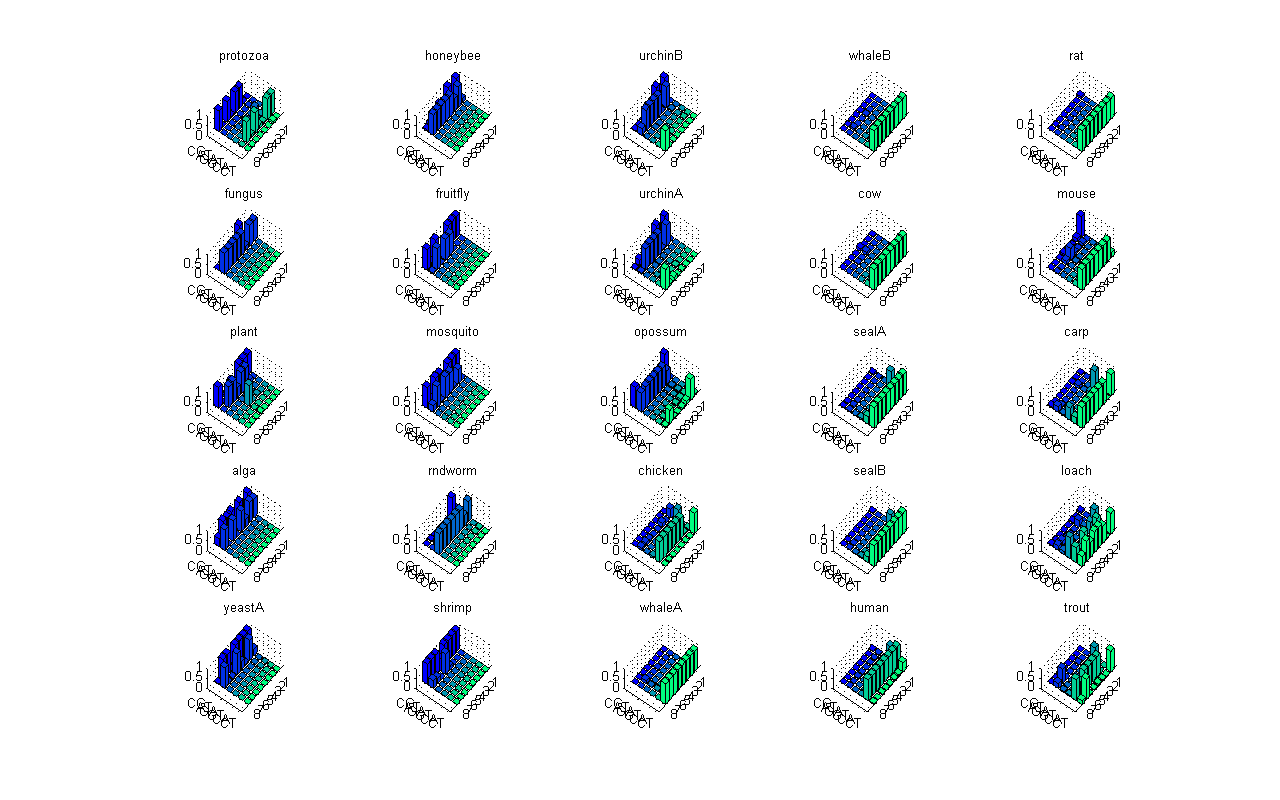}
	\caption{eighths}   
   \label{fig:3d8ths}
 \end{figure}   

	These figures suggest that this phenomenon remains quite consistent across mitochondrial genomes, and that the similarities between specific pairs of nucleotide frequencies occurs even in very short sequences of mtDNA.
	  Also, the green peaks in the $\mathscr{D}$-spectra of the opossum in fig. \ref{fig:3d8ths} indicate that its mtDNA \textit{does} in fact have some vertebrate-like $\mathscr{D}$-spectrum characteristics albeit not appearing to fit with the other vertebrates in previous figures.  These vertebrate-like features do not exist in any of the other non-vertebrates except for the protozoan (keep in mind that the heights of the green bars for the case of the protozoan are an artifact of the normalization procedure implemented in order to make more bars visible, and the blue bars do in fact dominate the green bars consistently in the unnormalized version of this plot which is available in the supplementary document).
	  
% I still have to check if that last part is true for the protozoan. I'm guessing that now.

	Another conceivable extension would involve adding knowledge of nucleotide frequency domination to the above analyses. The knowledge of the $\mathscr{D}$-spectra alone seem to give a good indication of whether or not an organism is a vertebrate, but in conjuction with knowing which sets of nucleotides dominate the other two, one may be able to narrow down the clades even further. A preliminary result is displayed in fig. \ref{fig:YWdominance} below, where a particular peak in each $\mathscr{D}$-spectrum is colored blue if the corresponding pair of nucleotides is more abundant within the organism's mitochondrial genome than the opposite pair of nucleotides, and green otherwise.

\begin{figure}[h!]
   \centering
\includegraphics[scale=0.4]{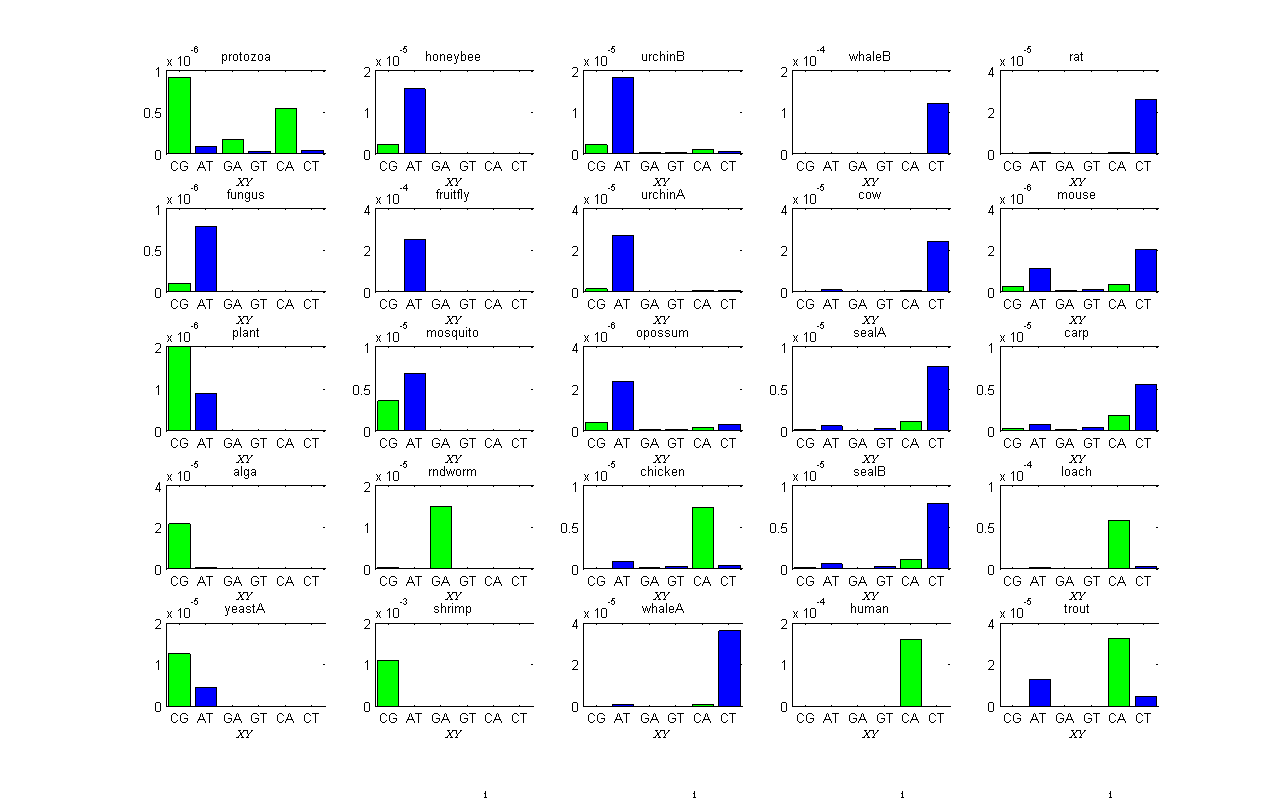}
	\caption{dominance}   
   \label{fig:YWDominance}
 \end{figure}   
	  
	  It is immediately apparent that whenever the first peak is at CG, C and G (green) happen to also be less common in the genome than A and T. Likewise, whenever the first peak is at AT it is blue (this includes the opossum), whenever the peak is at CA it is green and whenever it is at CT it is blue.
	  
\section{Open Problems}

A healthy number of open problems emerge from this study, both practical and theoretical. These problems span a wide range of areas ranging from probablistic combinatorics to physiology to theoretical chemical physics. From a practical standpoint, a method to optimize the above scheme without exhaustively searching each of 2$^{3969}-1$ possibilities could possibly improve the scheme presented in this paper (to more than 94.57\% accuracy). This could also make the process of generating such a scheme much more rigorous and therefore more practical. 

From a theoretical standpoint, one may wonder what biological factors lead to the sharp biases in nucleotide frequencies that allowed the above results to emerge. For example, significantly fewer first and second peaks arise at GA or GT in comparison with all other pairs of nucleotides. This is likely due to the fact that many organisms have a much lower guanine composition than the other three nucleotides, which is attributed to the reactive oxygen species (ROS). Likewise, one may wonder if there is some thermodynamic reason for these nucleotide biases. With recent developments in density functional theory, the electronic structure of short strands of nucleotides can be predicted \textit{ab initio}. If the potential energy surfaces for certain strands of nucleotides is found to be highly dependent on the nucleotide's environment, then perhaps the vertebrate environment allows certain configurations of DNA to be more thermodynamically favorable than others. Perhaps these configurations are not so favorable in other environments.

In addition to looking at the first and second peaks, one may investigate the third, and higher order peaks. Likewise, instead of just looking at the peaks, one  may investigate the troughs. Or, perhaps a combination of peaks and troughs. Further, rather than using an even exponent in eq. \ref{eq:metric} which forces all peaks to be positive, an odd exponent can be used. This will not affect the relative sizes of the peaks, but it will provide another piece of information: Which nucleotide within the pair is more abundant.

Another question that arises is whether or not this phenomenon is unique to mtDNA or if it remains consistent in other nucleic acids such as chromosomal DNA.

\end{document}